# Unpredictable repeatability in molecular evolution

Suman G Das[a,1] and Joachim Krug [a]

[a] Institute for Biological Physics, University of Cologne, Zuelpicher Strasse 77, D-50937 Cologne, Germany

**The extent of parallel evolution at the genotypic level is quantitatively linked to the distribution of beneficial fitness effects (DBFE) of mutations. The standard view, based on light-tailed distributions (*i.e.* distributions with finite moments), is that the probability of parallel evolution in duplicate populations is inversely proportional to the number of available mutations, and moreover that the DBFE is sufficient to determine the probability when the number of available mutations is large. Here we show that when the DBFE is heavy-tailed, as found in several recent experiments, these expectations are defied. The probability of parallel evolution decays anomalously slowly in the number of mutations or even becomes independent of it, implying higher repeatability of evolution. At the same time, the probability of parallel evolution is *non-self-averaging*, that is, it does not converge to its mean value even when a large number of mutations are involved. This behavior arises because the evolutionary process is dominated by only a few mutations of high weight. Consequently, the probability varies widely across systems with the same DBFE. Contrary to the standard view, the DBFE is no longer sufficient to determine the extent of parallel evolution, making it much less predictable. We illustrate these ideas theoretically and through analysis of empirical data on antibiotic resistance evolution.**

| Parallel evolution | Population genetics | Distribution of fitness effects | Predictability of evolution | Antibiotic resistance

The repeatability and predictability of evolution are important questions in the field of evolutionary biology. In 1990, Stephen Jay Gould famously mused about "replaying life's tape" (1). In subsequent years, the topic of parallel evolution has become a major subject of empirical research (2–4), and theoretical questions concerning the probability of parallel evolution within the mathematical theory of population genetics have also attracted substantial attention (5–7). Here the questions are focused mostly on changes at the level of genetic sequences. According to a common definition (6, 7), parallel evolution is said to occur when the exact same mutation is substituted in replicate populations. It is in this strong sense that we shall use the term *parallel evolution* here. The computation of the probability of parallel evolution is often set in a simplified scenario (5–7) where an asexual population evolves by strong selection and weak mutation (SSWM), which applies for moderately large population size and low mutation rate (more details in SI). The evolutionary process starts with a homogeneous population and $n$ possible beneficial mutations that can occur. Denoting by $r_i$ the substitution rate of the $i$'th mutation, $i = 1, 2, \ldots, n$, the probability that the $i$-th mutation will be the first to fix is $W_i = r_i/(\sum_j r_j)$. Therefore the probability that $k$ replicate populations will all fix the same mutation is given by $P_k = \sum_i W_i^k$. Although we focus on the repeatability of the first substitution event, this measure can be generalised to address evolutionary trajectories with several mutations (see (2) and further discussion in SI.)

Within the SSWM approximation, the substitution rate of a mutation is the product of its mutation rate and fixation probability, $r_i = \mu_i \pi_i$, and the repeatability measure Eq. (1) is affected equally by heterogeneities in $\mu_i$ and $\pi_i$ (4). However, because empirical information on mutation rate heterogeneity is sparse, theoretical studies have commonly assumed that all mutations occur at the same rate. If in addition the selection coefficients are small, $s_i \ll 1$, it follows that $r_i \sim \pi_i \sim s_i$, and the probability of parallel evolution is given by

$$P_k = \sum_i \frac{s_i^k}{\left(\sum_j s_j\right)^k}. \qquad [1]$$

We will adopt this simplification throughout, but emphasize that our theoretical results hold equally well for the full expression of $P_k$ with $r_i$ in place of $s_i$, provided the distribution of the substitution rates $r_i$ is heavy-tailed in the sense specified below.

The selection coefficients follow a distribution denoted by $P_s(s)$, which we refer to as the distribution of beneficial fitness effects (DBFE) (8). The mean probability of parallel evolution is $\langle P_k \rangle$ where $\langle \cdot \rangle$ denotes the average with respect to the DBFE. It has been hypothesized (5–7) that, since viable organisms are already relatively well-adapted, mutants of higher fitness must be chosen from the tails of fitness distributions, and therefore the DBFE $P_s(s)$ must have the form of a limiting distribution in extreme-value theory (EVT). The EVT hypothesis predicts that the DBFE belongs to one of three classes of distributions: the Weibull and Gumbel classes contain distributions with finite moments, whereas the Fréchet class contains distributions with power law tails (and therefore diverging moments). In the last case, the asymptotic form of the DBFE is

$$P_s(s) \sim \frac{A}{s^{1+\alpha}} \qquad [2]$$

with a scale parameter $A > 0$ and the tail exponent $\alpha > 0$. The cases of the Gumbel and Weibull extreme-value distributions have been explored in some detail for $k = 2$ (5–7). The Fréchet EVT class had been conjectured to be relatively unimportant biologically (7), but several subsequent studies (9–11) have uncovered signatures of heavy-tailed distributions of fitness effects (see SI for further details on tails of empirical DBFEs). In the realistic case of a large number of available beneficial mutations $n$, the statistics of $P_k$ for heavy-tailed distributions of the form Eq. (2) is markedly different from that of light-tailed distributions, as will be shown below.

**Results.** The number $n$ of beneficial mutations that can occur in a population varies widely across organisms and environments, but it is likely to be large. For bacterial populations, one may conservatively estimate that there are several thousand beneficial mutations (see SI). We therefore focus on $P_k$ in the large-$n$ regime. The simplest computation is in the limiting case of neutral variation where all selection coefficients are identical. Since all mutations are equally likely to be the first to fix, $P_k = 1/n^{k-1}$ (which is exact for all $n$). Specifically, the probability of parallel evolution in two replicates is $P_2 = 1/n$. Using this observation, one can define, for any system, the quantity $n_e = P_2^{-1}$ as the *effective number of mutations* that contribute to parallel evolution. It can be interpreted as the number of mutations in a different system which has the same $P_2$ but where all the mutations are equally likely to fix in the population. Therefore $n_e$ is a measure of the number of mutations that dominate the dynamics of fixation. It is similar to the notion of the effective number of reproducing lineages studied in (12) in the context of family size distributions.

The most commonly studied class of DBFEs is where all the moments are finite. In this case, the numerator and denominator in each term in Eq. (1) become uncorrelated as $n \to \infty$ and the distribution of $(\sum_i s_i^m)/n$ becomes sharply centred around the moment $\langle s^m \rangle$ of $P_s(\cdot)$. Therefore, for large $n$, we have

$$P_k \simeq \frac{\langle s^k \rangle}{\langle s \rangle^k} n^{-(k-1)}. \qquad [3]$$

The solid brown line in Fig 1(a) shows this behavior of $P_k$ for the exponential distribution. Notice that so far we have omitted the angular brackets around $P_k$, since it converges to the mean value in the limit of large $n$, as shown by the highly localized distribution of $P_k$ in Fig 1(b) (red dashed curve). Such quantities are described as *self-averaging*. For the particular case $k = 2$, we have $P_2 \sim \frac{1}{n}$, which is the characteristic decay in self-averaging systems. It was shown in (6) that for an exponential distribution, $\langle P_2 \rangle = 2/(n+1)$ (see SI for a general expression for $\langle P_k \rangle$). Our focus here, however, is on heavy-tailed distributions with tails of the form Eq. (2). When $k < \alpha$, Eq. (3) continues to hold. Particularly for $\alpha > 2$, $P_2$ still decays as $\sim 1/n$, as shown in Fig 1(a). However, when $k > \alpha$, the moment $\langle s^k \rangle$ diverges and Eq. (3) no longer holds. We can now break the analysis down into two cases.

Case I: The moderately heavy-tailed case occurs when $\alpha > 1$; in this case $\langle s \rangle$ is finite, but higher moments corresponding to $k > \alpha > 1$ diverge. For $k > \alpha$, the asymptotic behavior of $\langle P_k \rangle$ is

$$\langle P_k \rangle \simeq C_k n^{-(\alpha-1)}, \qquad [4]$$

where the constant $C_k = A\Gamma(k-\alpha)\Gamma(\alpha)/(\Gamma(k)\langle s \rangle^\alpha)$. Note that $\langle P_k \rangle$ decays with an exponent less than $k-1$; therefore the mean probability of parallel evolution is asymptotically much larger than in the case of light-tailed DBFEs. The scaling $n^{-(\alpha-1)}$ in Eq. (4) was first reported in (9) and recently derived independently in (12) in a different context. In particular, we see that when $1 < \alpha < 2$, $\langle P_2 \rangle$ decays anomalously, *i.e.* with an exponent $< 1$, in contrast to $P_2 \sim n^{-1}$ as in the light-tailed case; see Fig 1(a).

It is important to point out that $P_k$ does not become sharply centred around its mean value when $k > \alpha$, which can be shown as follows. The $m$-th moment is given by $\langle P_k^m \rangle = \langle P_{km} \rangle$ (see SI). The value of $\langle P_{km} \rangle$ can be read off from Eq. (4) by replacing $k$ by $km$. Thus, all moments are of the same order $n^{-(\alpha-1)}$. In particular, we notice that for $1 < \alpha < 2$, $\langle P_2^2 \rangle/\langle P_2 \rangle^2 \sim n^{\alpha-1}$. For self-averaging systems (which obey Eq. (3) for all $k$), this ratio goes asymptotically to 1, and the standard deviation vanishes relative to the mean (as seen in the red dashed curve in Fig 1(b)). In contrast, here we see that the standard deviation diverges relative to the mean, implying a broad distribution for $P_2$, as illustrated in Fig 1(b). This non-self averaging effect arises because the sum $P_k = \sum_i W_i^k$ is dominated by the largest weight $W_{\max}^k$ (12, 13). According to EVT, the largest selection coefficient scales as $n^{1/\alpha}$, implying that $W_{\max} \sim n^{1/\alpha - 1}$. Therefore, the scale of typical $P_k$ is

$$P_k \sim n^{k(\frac{1}{\alpha}-1)} \qquad [5]$$

for $k > \alpha$, which is asymptotically smaller than $\langle P_k \rangle$ as given by Eq. (4). In fact, most of the weight is concentrated near the typical value, and the much higher mean is obtained from values of $P_k$ that are much rarer but have much higher magnitude.

Case II: For the severely heavy-tailed case $0 < \alpha < 1$, all integer moments of $s$ diverge. It was shown in (13) that for a power-law distribution with $0 < \alpha < 1$,

$$\langle P_k \rangle \simeq \frac{\Gamma(k-\alpha)}{\Gamma(k)\Gamma(1-\alpha)} \qquad [6]$$

in the limit of large $n$. Specifically, the average probability of parallel evolution in two replicates is $\langle P_2 \rangle \simeq 1 - \alpha$. Note that the asymptotic form in Eq. (6) is independent of $n$, and thus we have the striking result that the probability of parallel evolution remains finite even in the limit of an infinite number of available alternative mutations. In the present case, all moments of $P_2$ are of $O(1)$, and therefore $P_2$ is non-self averaging. This is visible in the wide distribution of $P_2$ as shown in the numerically sampled plot in the inset of Fig 1(b). Similar non-self-averaging effects are familiar in the physics of disordered systems (see (13) and references therein), and in probability theory (14).

While the moderately ($\alpha > 1$) and severely ($\alpha < 1$) heavy-tailed cases display somewhat different behavior, we note that both Eq. (4) and Eq. (6) give rise to the recursion relation

$$\frac{\langle P_{k+1} \rangle}{\langle P_k \rangle} = 1 - \frac{\alpha}{k} \quad \text{for} \quad k \geq 2, \qquad [7]$$

which therefore holds for the entire range $0 < \alpha < k$. The result is independent of $n$ and of all features of the underlying distribution except the tail exponent $\alpha$. It is therefore suitable for extracting $\alpha$ from empirical data; however, the disadvantage is that the averages require large datasets. Equation (7) easily yields an approximate solution for large $k$, $\langle P_k \rangle \sim 1/k^\alpha$, which is verified in Fig 1(c). The slow decay of $\langle P_k \rangle$ with $k$ contrasts with the exponential decay of the *typical* $P_k$ as given by Eq. (5).

The theoretical results discussed so far are valid in the limit of large $n$. Nonetheless, we will show that signatures of non-self-averaging effects can be discovered in limited empirical data sets. For this purpose, we use data on selection coefficients associated with antibiotic resistance evolution reported in (9). In this study, the fitness effects of 48 beneficial mutations in the resistance enzyme TEM-1 $\beta$-lactamase were reported for *Escherichia coli* growing at four different concentrations of the antibiotic cefotaxime (see SI for further details of the experiment). An analysis based on EVT indicated that the DBFE is light-tailed for the lowest concentration and heavy-tailed for the three higher concentrations, although large uncertainties were associated with the exponents in the latter case (9). We evaluate the statistics of $P_2$ for the four different concentrations (see SI for further details). Fig 1(d) shows $P_2$ as a function of $n$. For the lowest concentration, $P_2$ is seen to be small with a small dispersion, and it decreases with $n$, consistent with our expectation. For the three higher concentrations, the values of $P_2$ are larger and have a large dispersion, which is consistent with heavy tailed distributions. There is no discernible decrease with $n$. However, due to the relatively small values of $n$ and the modest size of the data sets, it is not possible to distinguish this from a slow decrease with $n$. In Fig 1(e), we have plotted $P_k$ as a function of $k$. Note that the distinction between the typical and mean values (12) has important implications here. Due to the limited size of the data, we have not used the recursion relation Eq. (7) to infer $\alpha$. Instead, for each concentration, we have used the entire set of selection coefficients to create a single sample value of $P_k$, which is expected to be of the typical scale given by Eq. (5). Using this, we estimate the exponent $\alpha$ which is seen to progressively decrease with increasing concentration. indicating an increasingly heavy-tailed distribution. Thus, stronger selection pressures amplify the differences between fitness effects of beneficial mutations, leading to a broader distribution. Nonetheless, we should mention that inferred power-law exponents should be treated with some caution, since these can be sensitive to experimental errors or methods of analysis (9, 11). What is clear from Fig 1, however, is that the behavior of $P_k$ is at least a good qualitative indicator of the dispersion of selection coefficients. We have also computed and plotted the effective mutation number $n_e$ in Fig 1(f). The trend is again seen to be as predicted by theory. At the lowest concentration, $n_e$ is relatively large and close to $(n+1)/2$ (where $n$ is the actual number of mutations), consistent with an exponential distribution of selection coefficients. The effective mutation number decreases progressively with increasing concentration, and indicates a slower than exponential tail.

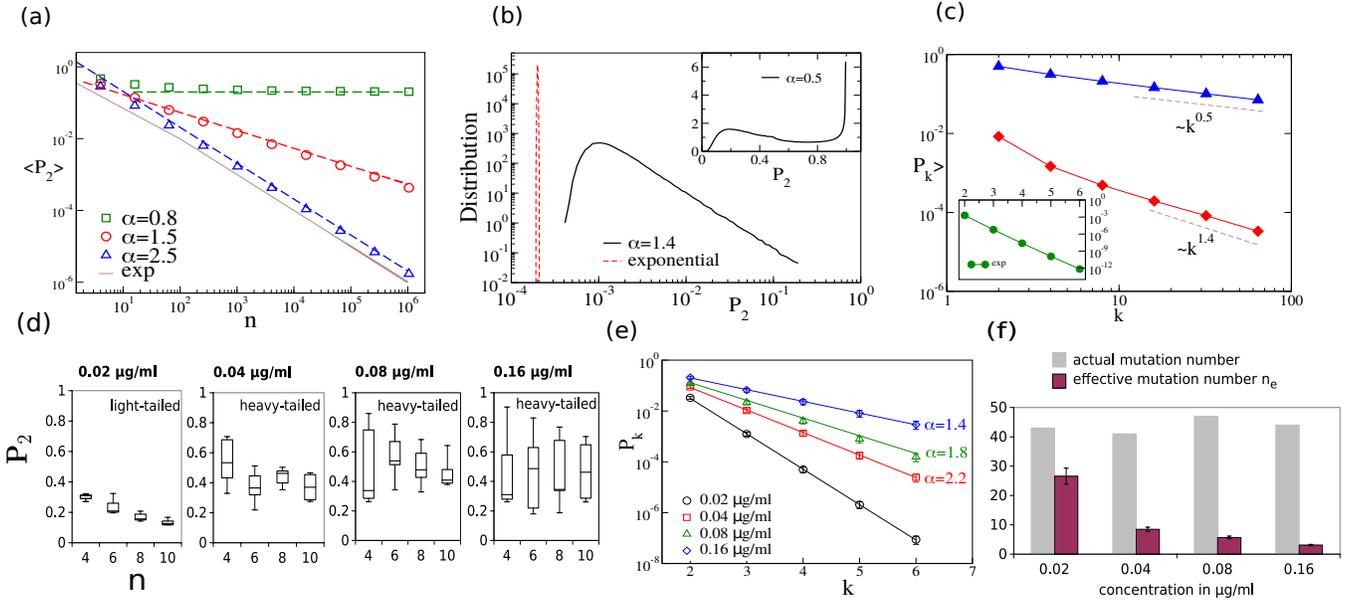

**Fig. 1.** (a) Plot of $\langle P_2 \rangle$ for three different values of $\alpha$. The symbols are numerically generated data with averages over $10^5$ realizations, and the dashed lines are the theoretical predictions from Eq. (4) and Eq. (6). The solid brown curve is the exact result for the exponential distribution. The curve for $\alpha = 2.5$ has the same asymptotic behavior $P_2 \sim n^{-1}$ as that of the exponential since $\alpha > k = 2$. (b) The black curve is the numerically sampled distributions of $P_2$ for $\alpha = 1.4$ and the inset shows the same for $\alpha = 0.5$; we used $10^6$ realizations and $n = 10^4$ mutations. The dashed red curve is the distribution of $P_2$ for an exponential distribution of selection coefficients and $n = 10^4$. (c) $\langle P_k \rangle$ as a function of $k$, from Eq. (4) and Eq. (6). We used $n = 10^4$ and $10^5$ realizations. The inset shows the rapid decay of $\langle P_k \rangle$ for an exponential distribution and $n = 10^3$, plotted using the exact result in SI. (d) This and the following figures analyze data from the study based on mutant screening reported in (9), which determined the selection coefficients for several resistance-conferring mutations in TEM-1 $\beta$-lactamase. Here we numerically estimate the distribution of $P_2$ from the selection coefficients reported in (9). The data set at each cefotaxime concentration was randomly split into subsets of size $n$ in order to obtain distributions of $P_2$ as a function of $n$. The box plots show median, quartiles, and extreme values. (e) The $P_k$ were obtained from the entire data set at each concentration and Eq. (5) were used to infer $\alpha$. (f) The effective mutation number $n_e = 1/P_2$ has been computed and compared with the actual number of mutations in the available dataset at each concentration.

**Conclusions.** Parallel phenotypic evolution often proceeds through distinct genotypic pathways. Here we have shown that heavy-tailed DBFEs can substantially enhance the probability of parallel evolution even at the genotypic level. However, we also find that this probability varies widely across large, independent samples generated from the same heavy-tailed DBFE. This makes it harder to generalize the degree of repeatability from one model system to other, closely related ones. On the flip side, the evolutionary process is dominated by a few mutations of high weight, making evolution of the system more repeatable, and therefore more predictable. For example, in the study (15) on antibiotic resistance evolution, the authors found that the mutation of highest effect (which also features in the heavy-tailed distributions reported in (9) and discussed above) occurred in the majority of multiple replicate experiments. The full implications of these ideas in the context of natural populations remain to be elucidated.

SGD and JK acknowledge support by the Deutsche Forschungsgemeinschaft (DFG, German Research Foundation) within SFB 1310 *Predictability in evolution*. We thank Arjan de Visser for helpful comments on the manuscript.

1. Gould SJ (1990) *Wonderful life: the Burgess Shale and the nature of history*. (WW Norton & Company).
2. De Visser J, Krug J (2014) Empirical fitness landscapes and the predictability of evolution. *Nature Reviews Genetics* 15(7):480–490.
3. Storz JF (2016) Causes of molecular convergence and parallelism in protein evolution. *Nature Reviews Genetics* 17(4):239–250.
4. Bailey SF, Blanquart F, Bataillon T, Kassen R (2017) What drives parallel evolution? how population size and mutational variation contribute to repeated evolution. *BioEssays* 39(1):1–9.
5. Gillespie JH (1983) A simple stochastic gene substitution model. *Theoretical population biology* 23(2):202–215.
6. Orr HA (2005) The probability of parallel evolution. *Evolution* 59(1):216–220.
7. Joyce P, Rokyta DR, Beisel CJ, Orr HA (2008) A general extreme value theory model for the adaptation of DNA sequences under strong selection and weak mutation. *Genetics* 180(3):1627–1643.
8. Bataillon T, Bailey SF (2014) Effects of new mutations on fitness: insights from models and data. *Annals of the New York Academy of Sciences* 1320(1):76–92.
9. Schenk MF, Szendro IG, Krug J, de Visser JAG (2012) Quantifying the adaptive potential of an antibiotic resistance enzyme. *PLoS Genetics* 8(6):e1002783.
10. Bank C, Hietpas RT, Wong A, Bolon DN, Jensen JD (2014) A Bayesian MCMC approach to assess the complete distribution of fitness effects of new mutations: uncovering the potential for adaptive walks in challenging environments. *Genetics* 196(3):841–852.
11. Foll M, et al. (2014) Influenza virus drug resistance: a time-sampled population genetics perspective. *PLoS Genetics* 10(2):e1004185.
12. Niwa HS (2022) Reciprocal symmetry breaking in Pareto sampling. *arXiv preprint arXiv:2202.04865*.
13. Derrida B (1997) From random walks to spin glasses. *Physica D: Nonlinear Phenomena* 107(2-4):186–198.
14. Pitman J, Yor M (1997) The two-parameter Poisson-Dirichlet distribution derived from a stable subordinator. *The Annals of Probability* pp. 855–900.
15. Van Dijk T, Hwang S, Krug J, de Visser JAG, Zwart MP (2017) Mutation supply and the repeatability of selection for antibiotic resistance. *Physical Biology* 14(5):055005.

**Supporting Information Text**

**Probability of fixation of beneficial mutations**

The probability of fixation of the $i$-th mutation is proportional to $s_i$ when $s_i$ is positive but small. Denoting $F_i$ as the fitness of the $i-th$ mutant arising in the background genotype with fitness $F_0$, we have $s_i = \frac{F_i - F_0}{F_0}$ (by definition). We require this quantity to be small, even though the fitness effects $S_i \equiv F_i - F_0$ are chosen from a distribution supported over the half-line $[0, \infty)$, which means that $S_i$ can be arbitrarily high. Thus, for the small $s$ approximation to hold, $F_0$ must be large enough such that, with high probability, the largest of $n$ chosen values of $s$ is much smaller than $F_0$. Let the distribution of the largest value be $P_e(s; n)$, and we define $C_e(s; n) = \int_s^\infty dx\, P_e(x; n)$. We therefore require that $C_e(\epsilon F_0; n) \ll 1$, where $\epsilon \ll 1$. For power law tails of $P_s(s)$ as considered here, it is known that $C_e(\epsilon F_0; n) \sim \frac{n}{(\epsilon F_0)^\alpha}$, and therefore the approximation holds when $F_0 \gg \left(\frac{n}{\epsilon}\right)^{\frac{1}{\alpha}}$.

**Strong selection and weak mutation (SSWM) regime**

In the SSWM regime, the conditions $\mu N \ll 1 \ll sN$ must be obeyed, where $\mu$ is the mutation rate, $N$ is the population size, and $s$ is a typical selection coefficient (1, 2). These conditions ensure that the population remains homogeneous most of the time, since the time to fixation of a beneficial mutation is much smaller than the time interval between the occurrence of beneficial mutations. This regime requires very low mutation rates and is somewhat artificial, but becomes less so when the DBFE is broad. A sample of selection coefficients drawn from a broad distribution is dominated by few values, and therefore mutants reach fixation quickly. When $\mu N > 1$ multiple mutant clones are present simultaneously and compete for fixation. As a consequence the distribution of fixed mutations is biased towards mutations of large effects, and repeatability increases compared to the SSWM regime (3, 4).

**Repeatability of evolutionary paths**

In the main text we have considered the repeatability of fixation of a single mutation. The population genetics literature on the repeatability of evolutionary pathways comprising several steps is limited. However, the quantity

$$P_k = \sum_i W_i^k \quad [1]$$

introduced in the main text has also been used as a measure for path repeatability in (5, 6), where the $W_i$ represent weights of evolutionary paths; see also (4) for a study using a different repeatability measure. Within the SSWM regime, the weight of a path is given by the product of the relative fixation probabilities $r_i/(\sum_j r_j)$ for the mutational steps along the path, and is similarly affected by the DBFE.

**Tails of empirical DBFEs**

Empirical estimates of the tails of DBFEs are available for a number of different systems. Below we discuss reports in the literature of both light and heavy-tailed DBFEs.
***Light-tailed DBFEs:*** Several studies have found evidence of distributions with finite moments. For example, evidence of exponential distributions were found for *Pseudomonas fluorescens* (7) across a range of environments, and for *Pseudomonas aeruginosa* (8) at low antibiotic concentration. A normal distribution was reported (9) for *P. fluorescens* in an environment where the ancestral type has extremely low fitness. While the normal distribution is light-tailed, it is not a limiting EVT distribution (which is not particularly surprising here (9), since the EVT hypothesis assumes that the ancestral type is relatively well-adapted). The DBFE for two bacteriophage viruses have been found (10) to be consistent with the uniform distribution, which is a special case of the limiting Weibull distribution of EVT. The DBFE for a protein in the hepatitis C virus was found (11) to be consistent with an exponential distribution. A study (12) on the protein Hsp90 in the yeast *Saccharomyces cerevisiae* determined the DBFE across several environments and estimated it to be of the Weibull type in all but one environment.
***Heavy-tailed DBFEs:*** A number of experiments have uncovered evidence for heavy-tailed distributions as well. For example, a study with an antibiotic resistance enzyme in *E. coli* (13) found evidence of heavy-tailed distributions at high antibiotic concentrations, with $\alpha$ belonging to both the moderately ($\alpha > 1$) and severely ($\alpha < 1$) heavy-tailed regimes. At low antibiotic concentration, the same study estimated the DBFE to be light-tailed. This is consistent with the previously mentioned work with *P. aeruginosa* (8) which reported that, at high drug concentration (and therefore low wild type fitness), the DBFE was broader and could not be fit with an exponential distribution. A study (14) on the influenza A H1N1 virus inferred that the DBFE belongs to the Weibull domain in the absence of the antiviral drug oseltamivir and to the Fréchet domain in its presence. The previously discussed work on Hsp90 in yeast (12) found the fitness effects to be heavy-tailed, with $\alpha > 2$, in an environment with lowered temperature and elevated salinity. in a related study (15), the DBFE of synonymous mutations in Hsp90 was found to be heavy-tailed in several environments with $\alpha > 2$ in most cases. A heavy-tailed DBFE has also been detected among mutations in tumors (16). A common pattern in many of these results is that the DBFE becomes broader in novel and challenging environments.

## Number of available beneficial mutations

The fraction of all mutations (occurring before the biasing effect of selection) that are beneficial is expected to vary strongly across organisms and environments, and is difficult to determine. Nevertheless, studies based on laboratory experiments with microbes have often concluded that beneficial mutations are relatively common. For example, Schenk et al. (13) estimate that 3.4 % of all base pair substitutions in the TEM-1 $\beta$-lactamase gene increase resistance to cefotaxime, and two survey articles (3, 17) state that the typical beneficial fraction among genome-wide mutations in bacteria is on the order of $10^{-2}$. The *E. coli* genome has about $4.6 \times 10^6$ base pairs (18). Therefore, even if one considers only point mutations, there are potentially about $10^4$ beneficial mutations.

## Probability of parallel evolution for exponential distributions

Following (19), we notice that, by definition,

$$\Gamma(k) = (\sum_j s_j)^k \int_0^\infty dx \ e^{-(\sum_j s_j)x} x^{k-1}, \qquad [2]$$

and therefore we have

$$P_k = \int_0^\infty \frac{dx \ x^{k-1} e^{-x \sum_j s_j}}{\Gamma(k)} \sum_i s_i^k. \qquad [3]$$

The mean probability is

$$\langle P_k \rangle = \frac{n}{\Gamma(k)} \int_0^\infty dx \ x^{k-1} \langle e^{-xs} \rangle^{n-1} \langle s^k e^{-xs} \rangle. \qquad [4]$$

Note that one can simply use the exponential distribution $P_s(s) = e^{-s}$, since a rescaling of the distribution does not alter the homogeneous expression for $P_k$. It is easy to show that for this distribution, $\langle e^{-xs} \rangle = 1/(1+x)$ and $\langle s^k e^{-xs} \rangle = \Gamma(k+1)/(1+x)^{k+1}$. Substituting these in equation Eq. (4) we get

$$\begin{aligned}\langle P_k \rangle &= kn \int_0^\infty dy \ \frac{y^{k-1}}{(1+y)^{n+k}} \\ &= kn \int_0^\infty dy \ \frac{\left((1+y)-1\right)^{k-1}}{(1+y)^{n+k}}.\end{aligned} \qquad [5]$$

Performing the binomial expansion of $\left((1+y)-1\right)^{k-1}$ and evaluating the subsequent integrals leads to

$$\langle P_k \rangle = nk \sum_{m=0}^{k-1} \binom{k-1}{m} \frac{(-1)^{k-1-m}}{(k-1)+n-m}, \qquad [6]$$

which reduces to the result of Orr (20) for $k = 2$.

## Mean probability of parallel evolution for moderately heavy-tailed distributions

Here we obtain the asymptotic expression of $\langle P_k \rangle$ for $k > \alpha > 1$. A closely related proof is available in (21). For large $n$, the integral in Eq. (4) is dominated by the small $x$ behavior, and for sufficiently small $x$ we have

$$\begin{aligned}\langle e^{-xs} \rangle &= \int e^{-xs} P_s(s) ds \simeq \int (1-sx) P_s(s) ds & [7] \\ &= 1 - x\langle s \rangle \simeq e^{-x\langle s \rangle}. & [8]\end{aligned}$$

Further, for $k > \alpha$ (19),

$$\langle s^k e^{-xs} \rangle \simeq \int s^k e^{-xs} \frac{A}{s^{1+\alpha}} = A x^{\alpha-k} \Gamma(k-\alpha), \qquad [9]$$

where the first step is justified because, for sufficiently small $x$, the dominant contribution to the integral comes from the tail of the distribution. We now use Eq. (8) and Eq. (9) in Eq. (4) to obtain the result in the main text:

$$\langle P_k \rangle = A \frac{\Gamma(k-\alpha)\Gamma(\alpha)}{\Gamma(k)\langle s \rangle^\alpha} n^{-(\alpha-1)}.$$

## Moments of the probability of parallel evolution for moderately heavy-tailed distributions

The $m$-th moment of $P_k$ is given by

$$\langle P_k^m \rangle = \sum_i \langle W_i^{km} \rangle + \text{off-diagonal terms} \quad [10]$$

$$\simeq \sum_i \langle W_i^{km} \rangle$$

$$= \langle P_{km} \rangle$$

$$\simeq \frac{A}{\langle s \rangle^\alpha} \frac{\Gamma(mk-\alpha)\Gamma(\alpha)}{\Gamma(mk)} n^{-(\alpha-1)}, \quad [11]$$

where the second term in Eq. (10) is dropped because, as we show now, it is of sub-leading order. The off-diagonal terms in $(\sum_{i=1}^n W_i)^m$ can be written as $\sum_{j=1}^{m-1} C_j$ where

$$C_j = \sum_{\{i_p\}} W_{i_1}^j \prod_{p=2}^{m-j+1} W_{i_p}. \quad [12]$$

The sum has $\binom{n}{m-j+1}$ terms, which, in the limit of large $n$, is $O(n^{m-j+1})$ terms. Using the form Eq. (2) of the $\Gamma$-function, we write

$$\langle W_{i_1}^j \prod_{p=2}^{m-j+1} W_{i_p} \rangle$$

$$= \frac{1}{\Gamma(km)} \int dx\ e^{-x(\sum_{i=1}^n s_i)} x^{km-1} s_{i_1}^{jk} \prod_{p=2}^{m-j} s_{i_p}^k$$

$$= \frac{1}{\Gamma(km)} \int dx\ \langle e^{-xs(n-m)} \rangle x^{km-1} \langle s^{jk} e^{-xs} \rangle \langle s^k e^{-xs} \rangle^{m-j}$$

$$\simeq C \int dx\ e^{-x(n-m)\langle s \rangle} x^{km-1} x^{\alpha-jk} x^{(\alpha-k)(m-j)}$$

$$= C(n-m)^{(\alpha)(j-m-1)} \int dy\ e^{-y\langle s \rangle} y^{\alpha(1+m-j)-1}$$

where $C = A^{1+m-j}\Gamma(jk-\alpha)(\Gamma(k-\alpha))^{m-j}/\Gamma(km)$. Note that the $n$-dependence is $n^{-\alpha(m-j+1)}$, and since there are $O(n^{m-j+1})$ such terms, $C(j) \sim n^{-(\alpha-1)(1+m-j)}$. Since $j \leq m-1$, the exponent $(\alpha-1)(1+m-j) > \alpha-1$, and therefore $C_j$ decays faster than the leading term given in Eq. (11).

## Numerical simulations

For Figs 1(a)-(c) of the main text, large random samples were generated from heavy-tailed distributions. We used a Monte Carlo procedure outlined in (19). We recapitulate it here for completeness. First, we generate independent random numbers $\{x_1, x_2, \ldots\}$ distributed uniformly between 0 and 1. Then we write $s_1 = (-\ln x_1)^{-\frac{1}{\alpha}}$, and use the recursion relation

$$s_{i+1} = s_i(1 - s_i^\alpha \ln x_{i+1})^{-\frac{1}{\alpha}} \quad [13]$$

to generate the sequence $\{s_1, s_2, \ldots\}$. The value of $P_k$ calculated from this sequence has the desired statistics.

## Details of Data Analysis

The analyzed data was obtained from the study reported in (13).

## Brief description of the relevant experiments in Ref. (13)

In this study, PCR mutagenesis was used to introduce random mutations into TEM-1 $\beta$-lactamase on plasmids in *Escherichia coli*. 48 point mutations with beneficial effects on cefotaxime (CTX) resistance were identified by mutant screening. Fitness at various cefotaxime concentrations were inferred from survival data using a branching model. Selection coefficients were computed relative to the least fit mutant at each CTX concentration.

**Data Analysis**

We have used the selection coefficients from (13) in our study. The expression for $P_k$ used for analysis was

$$P_k = \sum_i \frac{s_i^k}{(\sum_j s_j)^k}. \qquad [14]$$

In Fig 1B of the main text, we split the total dataset of the selection coefficients randomly into disjoint subsets at each concentration. Each subset was used to produce one value of $P_2$, and the boxplots were created from the values of $P_2$ obtained in this way. For Fig 1C, the entire dataset at each concentration was used to obtain the value of $P_k$. Error estimates were not available for the selection coefficients from (13). However, measurements of resistance level to cefotaxime in (13) had a mean error level of approximately 10 per cent which was used as a rough estimate of the error level in the selection coefficients. The dominant contribution to the variation in $P_k$ comes from the numerator of Eq. (14) and this was taken to be the only source of error as an approximation. Subsequently, standard error propagation was used to estimate the error in $P_k$ from that in the selection coefficients. The values of $\alpha$ were obtained by non-linear least squares fitting of the data. In Fig 1D, the actual mutation number was reported as the number of selection coefficients used for our analysis. The error bars in the effective mutation number were obtained through error propagation from $P_2$.

**References**

1. Gillespie JH (1983) A simple stochastic gene substitution model. *Theoretical population biology* 23(2):202–215.
2. Orr HA (2002) The population genetics of adaptation: the adaptation of DNA sequences. *Evolution* 56(7):1317–1330.
3. Sniegowski PD, Gerrish PJ (2010) Beneficial mutations and the dynamics of adaptation in asexual populations. *Phil Trans R Soc B* 365:1255–1263.
4. Szendro IG, Franke J, de Visser JAGM, Krug J (2013) Predictability of evolution depends nonmonotonically on population size. *Proc Natl Acad Sci USA* 110:571–576.
5. Roy SW (2009) Probing evolutionary repeatability: neutral and double changes and the predictability of evolutionary adaptation. *PLoS One* 4(2):e4500.
6. De Visser J, Krug J (2014) Empirical fitness landscapes and the predictability of evolution. *Nature Reviews Genetics* 15(7):480–490.
7. Kassen R, Bataillon T (2006) Distribution of fitness effects among beneficial mutations before selection in experimental populations of bacteria. *Nature Genetics* 38(4):484–488.
8. MacLean RC, Buckling A (2009) The distribution of fitness effects of beneficial mutations in Pseudomonas aeruginosa. *PLoS Genetics* 5(3):e1000406.
9. McDonald MJ, Cooper TF, Beaumont HJ, Rainey PB (2011) The distribution of fitness effects of new beneficial mutations in Pseudomonas fluorescens. *Biology letters* 7(1):98–100.
10. Rokyta DR, et al. (2008) Beneficial fitness effects are not exponential for two viruses. *Journal of Molecular Evolution* 67(4):368.
11. Dai L, et al. (2021) Quantifying the evolutionary constraints and potential of Hepatitis C virus NS5A protein. *mSystems* 6(2):e01111–20.
12. Bank C, Hietpas RT, Wong A, Bolon DN, Jensen JD (2014) A Bayesian MCMC approach to assess the complete distribution of fitness effects of new mutations: uncovering the potential for adaptive walks in challenging environments. *Genetics* 196(3):841–852.
13. Schenk MF, Szendro IG, Krug J, de Visser JAG (2012) Quantifying the adaptive potential of an antibiotic resistance enzyme. *PLoS Genetics* 8(6):e1002783.
14. Foll M, et al. (2014) Influenza virus drug resistance: a time-sampled population genetics perspective. *PLoS Genetics* 10(2):e1004185.
15. Fragata I, et al. (2018) The fitness landscape of the codon space across environments. *Heredity* 121(5):422–437.
16. Tokutomi N, Nakai K, Sugano S (2021) Extreme value theory as a framework for understanding mutation frequency distribution in cancer genomes. *PloS one* 16(8):e0243595.
17. Gordo I, Perfeito L, Sousa A (2011) Fitness effects of mutations in bacteria. *J Mol Microbiol Biotechnol* 21:20–35.
18. Blattner FR, et al. (1997) The complete genome sequence of escherichia coli k-12. *science* 277(5331):1453–1462.
19. Derrida B (1997) From random walks to spin glasses. *Physica D: Nonlinear Phenomena* 107(2-4):186–198.
20. Orr HA (2005) The probability of parallel evolution. *Evolution* 59(1):216–220.
21. Niwa HS (2022) Reciprocal symmetry breaking in Pareto sampling. *arXiv preprint arXiv:2202.04865*.